\documentclass{article}
\usepackage{arxiv}

\usepackage[utf8]{inputenc}
\usepackage[T1]{fontenc}    
\usepackage{hyperref}       
\usepackage{url}            
\usepackage{booktabs}       
\usepackage{amsfonts}       
\usepackage{nicefrac}       
\usepackage{microtype}      
\usepackage{graphicx}
\usepackage{xcolor}
\usepackage{todonotes}

\usepackage{float}
\floatstyle{boxed} 
\restylefloat{figure}

\usepackage{natbib}
\usepackage{doi}

\title{Architecture for a multilingual Wikipedia}
\author{Denny Vrande\v{c}i\'{c} \\ Google \\ vrandecic@google.com}
\date{Version 1 - April 1st, 2020}

\newcommand{\wikidata}[1]{\href{https://www.wikidata.org/entity/#1}{\underline{\textcolor{blue}{\texttt{#1}}}}}

\begin{document}

\maketitle

\begin{abstract}
Wikipedia's vision is a world in which everyone can share in the sum of all knowledge.
In its first two decades, this vision has been very unevenly achieved.
One of the largest hindrances is the sheer number of languages Wikipedia needs to cover in order to achieve that goal.
We argue that we need a new approach to tackle this problem more effectively,
a multilingual Wikipedia where content can be shared between language editions.
This paper proposes an architecture for a system that fulfills this goal.
It separates the goal in two parts:
creating and maintaining content in an abstract notation within a project called Abstract Wikipedia,
and creating an infrastructure called Wikilambda that can translate this notation to natural language.
Both parts are fully owned and maintained by the community,
as is the integration of the results in the existing Wikipedia editions.
This architecture will make more encyclopedic content available to more people in their own language,
and at the same time allow more people to contribute knowledge
and reach more people with their contributions,
no matter what their respective language backgrounds.
Additionally, Wikilambda will unlock a new type of knowledge asset
people can share in through the Wikimedia projects,
functions,
which will vastly expand what people can do with knowledge from Wikimedia,
and provide a new venue to collaborate and to engage
the creativity of contributors from all around the world.
These two projects will considerably expand the capabilities of the Wikimedia platform
to enable every single human being to freely share in the sum of all knowledge.
\end{abstract}

\fcolorbox{red}{white}{\begin{minipage}{0.978\linewidth}\textit{This
paper is work in progress.
Comments and suggestions for improvement are welcome and
it is expected that the paper will be updated following your comments.
Encouragements and substantiated criticisms are also welcome.
}\end{minipage}}

\section{Introduction}

In this paper we propose and describe two new projects, Abstract Wikipedia and
Wikilambda.\footnote{Note that both these names are just preliminary monickers,
and not the final names for the projects. As with everything in this proposal,
the eventual names and ultimate scope of these projects
are expected to emerge from a community process.}
Both are motivated by Wikimedia's vision to work toward
\textit{a world in which every single human being can freely share
in the sum of all knowledge}~\cite{vision},
and both have the potential to further democratize access to and participation in the creation of knowledge.

\textbf{Abstract Wikipedia} is a project to capture encyclopedic content
in a way that abstracts from natural language.
The goal is to allow everyone to contribute and read content in Abstract Wikipedia,
no matter what languages they understand,
and to allow people with different languages to work on and maintain the same content.
The abstract content gets rendered into a natural language for consumption.
Such a translation to natural languages is achieved through
the encoding of a lot of linguistic knowledge and of algorithms and
functions to support the creation of the human-readable renderings of the content.
These functions are captured in Wikilambda -- and this is why Abstract Wikipedia is built on Wikilambda.

\textbf{Wikilambda} is a wiki of functions. Functions take some input and return an output.
For Abstract Wikipedia, we will need functions that take abstract content as the input
and return natural language text as the output.
But the creation of such functions is not trivial, and requires
a rich environment to create and maintain functions.
Collaborative development is a dynamic field of research and development.
Wikilambda aims at making access to functions and writing and
contributing new functions as simple as contributing to Wikipedia,
thus aiming to create a comprehensive repository of algorithms and functions.

These two projects will considerably expand the capabilities of the Wikimedia platform
to enable every single human being to freely share share in the sum of all knowledge.
Abstract Wikipedia will make more encyclopedic content available to more people in their own language,
and at the same time allow more people to contribute knowledge
and reach more people with their contributions,
no matter what their respective language backgrounds.
Wikilambda will unlock a new type of knowledge asset people can share in through the Wikimedia projects,
functions~\cite{function}, which will vastly expand what people can do with knowledge from Wikimedia,
and provide a new venue to collaborate and engage the creativity of contributors from all around the world.

This text is structured as follows.
In Section~\ref{sec:motivation}, we discuss why a multilingual Wikipedia would be valuable,
and then provide a simple example of how the system works in Section~\ref{sec:example},
introducing terminology that we are going to use throughout.
We then spell out some requirements and challenges for the project in Section~\ref{sec:desiderata},
before describing the system as a whole in Section~\ref{sec:architecture}.
This leads to a discussion of Wikilambda and its wider impact in Section~\ref{sec:wikilambda}.
We turn our attention to a list of the biggest risks and open questions in Section~\ref{sec:risks}
but then in Section~\ref{sec:advantages} also to the unique advantages
that make the project more likely to succeed,
and in Section~\ref{sec:open} the numerous future research questions opened by this proposal.
We close with a discussion of alternative approaches in Section~\ref{sec:alternatives},
before summarizing the proposal in Section~\ref{sec:summary}.

\section{Motivation}
\label{sec:motivation}

Wikipedia is often described as a wonder of the modern age.
There are more than 50 million articles in about 300 languages.
The goal of allowing everyone to share in the sum of all knowledge is achieved, right?

Not yet.

The knowledge in Wikipedia is unevenly distributed.
Let's take a look at where the first twenty years of editing Wikipedia have taken us.
All language editions of Wikipedia are independently written and maintained.
The number of articles varies between the different language editions of Wikipedia:
English, the largest edition, has more than six million articles;
Cebuano, a language spoken in the Philippines, has 5.3 million articles;
Swedish has 3.7 million articles; and German has 2.4 million articles.\footnote{Cebuano
and Swedish have a large number of machine generated articles.}
In fact, the top nine languages alone hold more than half of all articles
across the Wikipedia language editions --- and if you take the bottom half of all Wikipedias ranked by size,
they together wouldn't have 10\% of the number of articles in the English Wikipedia.

It is not just the sheer number of articles that differs between editions, but their comprehensiveness does as well:
the English Wikipedia article on Frankfurt has a length of 184,686 characters,
a table of contents spanning 87 sections and subsections, 95 images, tables and graphs, and 92 references
--- whereas the Hausa Wikipedia article states that it is a city in the German state of Hesse,
and lists its population and mayor.
Hausa is a language spoken natively by forty million people and a second language to another twenty million.

It is not always the case that the large Wikipedia language editions have more content on a topic.
Although readers often consider large Wikipedias to be more comprehensive,
local Wikipedias may frequently have more content on topics of local interest:
all the English Wikipedia knows about the Port of C\u{a}l\u{a}ra\textcommabelow{s}i is that it is
one of the largest Romanian river ports,
located at the Danube near the town of C\u{a}l\u{a}ra\textcommabelow{s}i.
The Romanian Wikipedia on the other hand offers several paragraphs of content about the port.

The topics covered by the different Wikipedias also overlap less than one would initially assume.
English Wikipedia has six million articles, German has 2.4 million articles
--- but only 1.2 million topics are covered by both Wikipedias.
A full 1.2 million topics have an article in German --- but not in English.
The top ten Wikipedias by activity, each of them with more than a million articles,
have articles on only one hundred thousand topics in common.
In total, the different language Wikipedias cover 18 million different topics in over 50 million articles
--- and English only covers 31\% of the topics.

Besides coverage, there is also the question of how up to date the different language editions are:
in June 2018, San~Francisco elected London Breed as its new mayor.
Nine months later, in March 2019, we conducted an analysis of
who the mayor of San Francisco was stated to be,
according to the different language versions of Wikipedia~\cite{wp20vrandecic}.
Of the 292 language editions, a full 165 had a Wikipedia article on San Francisco.
Of these, 86 named the mayor.
The good news is that not a single Wikipedia lists a person who never was mayor of San Francisco
--- but the vast majority are out of date.
English switched the minute London Breed was sworn in.
But 62 Wikipedia language editions list an out-of-date mayor
--- and not just the previous mayor Ed Lee, who became mayor in 2011,
but also often Gavin Newsom (mayor from 2004 to 2011),
and his predecessor, Willie Brown (mayor from 1996 to 2004).
The most out-of-date entry is to be found in the Cebuano Wikipedia,
who names Dianne Feinstein as the mayor of San Francisco.
She had that role after the assassination of Harvey Milk and George Moscone in 1978,
and remained in that position for a decade until 1988 --- Cebuano was more than thirty years out of date.
Only 24 language editions had listed the current mayor, London Breed,
out of the 86 who listed the name at all.
Of these 24, eight were current because they used Wikidata~\cite{wikidata} to update itself automatically.

An even more important metric for the success of a Wikipedia are the number of contributors:
English has more than 31,000 active contributors\footnote{An
active contributor is a contributor with five or more edits within the last thirty days.}
--- three out of seven active Wikimedians are active on the English Wikipedia.
German, the second most active Wikipedia community, already only has 5,500 active contributors.
Only eleven language editions have more than a thousand active contributors
--- and more than half of all Wikipedias have fewer than ten active contributors.
To assume that fewer than ten active contributors can write and maintain a comprehensive encyclopedia
in their spare time is optimistic at best.
These numbers basically doom the vision of the Wikimedia movement to realize a world where everyone
can share in the sum of all knowledge.

Abstract Wikipedia aims to create and maintain the content of Wikipedia for all languages only once,
and then render it in the natural language of the reader.
This means coverage and currency would align much more between the articles,
and more knowledge would be available to more readers provided by more contributors.

A major change Abstract Wikipedia would bring along would be to change
the incentive infrastructure of Wikipedia~\cite{wikinomics,shirky}.
Today, contributors to less active language communities may be demotivated by the gargantuan
task of creating a comprehensive and current encyclopedia.
If, instead of having to create and maintain the content for their language, they `only' need to provide
the lexicographical knowledge and the renderings for their own language,
and can see large amounts of content being unlocked by their contributions,
they can gain new motivation for seeing a comprehensive and current encyclopedia in their language arise.
The impossibly daunting task becomes visibly possible.

Additionally, multilingual contributors do not have to choose which community to contribute to,
but can contribute to Abstract Wikipedia and thus make their knowledge available in many languages.
This increases their efficiency and their sense of satisfaction.
In the end, this will help with increasing the reach of underrepresented contributors by a large margin.

Finally, by enabling a global community we can reduce the risk of language editions
becoming ideologically biased~\cite{wp20vrandecic}.
This is a risk that is particularly prevalent if the number of contributors is small.
By having the content be created and maintained globally,
we can significantly reduce the risk of ideological lock-in
and ensure a larger diversity of points of views being represented,
thus leading to a higher quality of the resulting articles.

\section{Example}
\label{sec:example}

The following gives a toy example of how some text would be generated.
It obviously does not cover the whole complexity of the problem space,
but is meant to illustrate the system and to allow us
to introduce some terminology that we use throughout the paper.

Let us take the following two (simplified) sentences from English Wikipedia:

\textit{``San Francisco is the cultural, commercial, and financial center of Northern California.
It is the fourth-most populous city in California, after Los Angeles, San Diego and San Jose.''}

One advantage we have is that many of the entities in these two sentences are already represented as
unambiguous identifiers in Wikidata~\cite{wikidata}:
San Francisco is \wikidata{Q62}, Northern California \wikidata{Q1066807}, California \wikidata{Q99},
Los Angeles \wikidata{Q65}, San Diego \wikidata{Q16552}, and San Jose \wikidata{Q16553}.
Using identifiers instead of names we could thus rewrite the sentence to:

\textit{``\wikidata{Q62} is the cultural, commercial, and financial center of \wikidata{Q1066807}.
It is the fourth-most populous city in \wikidata{Q99},
after \wikidata{Q65}, \wikidata{Q16552}, and \wikidata{Q16553}.''}

These two sentences can be represented as abstract content as given in Figure~\ref{fig:example}.

\begin{figure}[tb!]
\centering
\begin{verbatim}
Article(
  content: [
    Instantiation(
      instance: San Francisco (Q62),
      class: Object_with_modifier_and_of(
               object: center,
               modifier: And_modifier(
                 conjuncts: [cultural, commercial, financial]
               ),
               of: Northern California (Q1066807)
             )
    ),
    Ranking(
      subject: San Francisco (Q62),
      rank: 4,
      object: city (Q515),
      by: population (Q1613416),
      local_constraint: California (Q99),
      after: [Los Angeles (Q65),
              San Diego (Q16552),
              San Jose (Q16553)]
    )
  ]
)
\end{verbatim}
\caption{An example content of two sentences describing San Francisco.}
\label{fig:example}
\end{figure}

The example shows a single \textbf{constructor} of type \texttt{Article}
with a single \textbf{key} called \texttt{content}.
The \textbf{value} of \texttt{content} is a \textbf{list} with two constructors,
one of type \texttt{Instantiation} the other of type \texttt{Ranking}.
The \texttt{Instantiation} constructor in turn has two keys, \texttt{instance} and \texttt{class},
where the \texttt{instance} key has a simple entity as the value,
\textit{San Francisco} (an \textbf{item} taken from Wikidata
and thus identified by \wikidata{Q62}),
and the \texttt{class} key has a \textbf{complex value} built from another constructor.

The names of the constructors and their keys and values are given here in English,
but that is merely convenience.
They would be given by internal identifiers that then in turn could have labels and descriptions
in any of the Wikipedia languages
(ideally, the labels and descriptions would in the end be represented by abstract text,
but we need to start with a simpler representation to get to the point where the system can support
such a form of inceptional self-hosting).

For each language, we then need a \textbf{renderer} for every type of constructor.
A simplified rough sketch for English is given in Figure~\ref{fig:render-en}
(entirely given in a convenient pseudo-syntax),
and for German in Figure~\ref{fig:render-de}.
The German renderer would lead to these two sentences:

\begin{figure}[tb!]
\centering
\begin{verbatim}
Instantiation(instance, class):
  instance + "is" + class
Object_with_modifier_and_of(modifier, object, of):
  (if modifier: modifier) + object + (if of: "of" + of)
Ranking(subject, rank, object, by, local_constraint, after):
  subject + "is" + "the" + ordinal(rank) + superlative(by) +
  object + (if local_constraint: "in" + local_constraint) +
  (if after: "," + "after" + after)
\end{verbatim}
\caption{Renderers for English for some of the constructors in Figure~\ref{fig:example}.}
\label{fig:render-en}
\end{figure}

\begin{figure}[tb!]
\centering
\begin{verbatim}
Instantiation(instance, class):
  instance + "ist" + class
Object_with_modifier_and_of(modifier, object, of):
  (if modifier: modifier) + object + (if of: genitive(of))
Ranking(subject, rank, object, by, local_constraint, after):
  subject + "ist" + (if after: "," + "nach" + after + ",") +
  article(definitive, gender_of(object)) + ordinal(rank) +
  superlative(by) + object +
  (if local_constraint: "in" + dative(local_constraint))
\end{verbatim}
\caption{Renderers for German for some of the constructors in Figure~\ref{fig:example}.}
\label{fig:render-de}
\end{figure}

\textit{``San Francisco ist das kulturelle, kommerzielle und finanzielle Zentrum Nordkaliforniens.
Es ist, nach Los Angeles, San Diego und San Jose, die viertgr\"{o}ßte Stadt in Kalifornien.''}

Note that in most cases, the renderers wouldn't actually result in strings
(as shown here for simplification),
but in objects such as noun phrases or clauses, which in turn have properties
and need renderers to be eventually turned into strings.

The renderers are all full-fledged \textbf{functions} which are created and maintained by the community.
As we can see in the sketched pseudo-syntax given here, they would in turn use other functions
from a large catalog of such functions, for example \texttt{gender\_of},
\texttt{dative}, \texttt{ordinal}, \texttt{if}, and \texttt{+}.

We glossed over many language specific issues like agreement~\cite{agreement,melcukaspects}, saliency~\cite{saliency}, and others.
For example, the hard-coded string \texttt{"is"} in the English \texttt{Instantiation} renderer could refer
to the appropriate \textbf{lexeme} in Wikidata~\cite{wikidatalex}
for the verb \textit{be}, \wikidata{L1883},
and then select the appropriate tense, person, and number,
i.e. the number and person would be based on the number and person of the \texttt{instance} value.

\section{Desiderata}
\label{sec:desiderata}

There are numerous requirements on the multilingual Wikipedia project.
We won't dwell on all of these --- e.g. the code must be open source,
the project must be run by the Wikimedia Foundation and thus needs to blend in to the current Wikimedia infrastructure,
all content must be available under an open license~\cite{open} compatible with Wikipedia, etc.
In this Section, we will discuss a few non-obvious requirements that constrain the possible solutions.

\begin{figure}[tb!]
\centering
\includegraphics[scale=0.6, trim=0 320 370 5, clip]{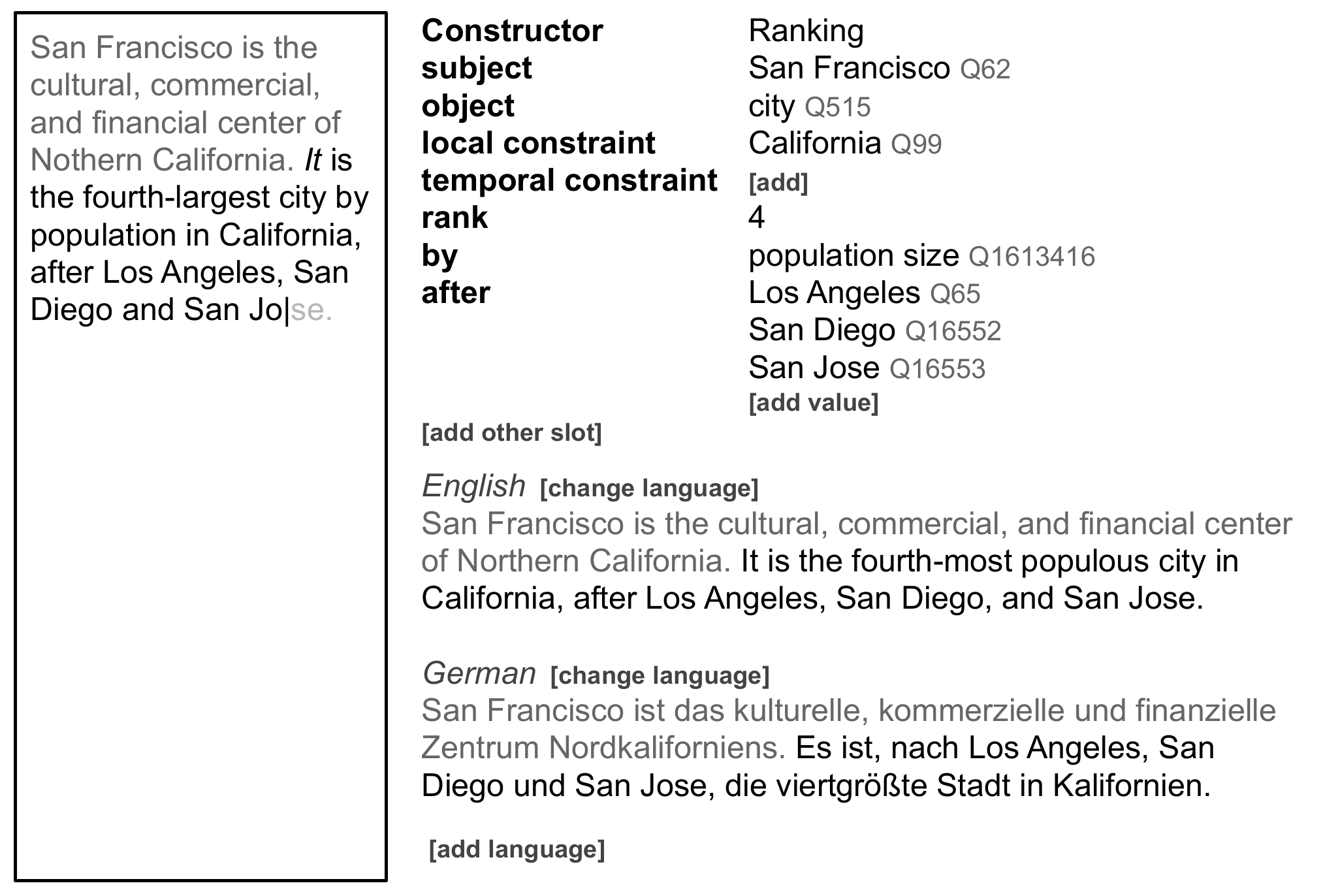}
\caption{
Mock up of the user interface.
On the left is a free text input box.
That text is classified in order to offer a first pass of the abstract content in the top right corner,
where the constructor and the respective values are being displayed and made available for direct editing.
The bottom right shows a rendering of the currently edited content in a selection of languages the contributor understand.
This gives a feedback loop that allows the contributor to adapt to the systems constructor library.
}
\label{fig:mockup}
\end{figure}

\paragraph{The content has to be editable in any of the supported languages.}
Note that this does not mean that we need a parser that can read arbitrary input in any language.
A very restricted editor, using forms, drop-down elements, and entity selectors,
would be sufficient and easy to internationalize and localize.
It would furthermore have the advantage that it would potentially work well with mobile interfaces.
A mock up is given in Figure~\ref{fig:mockup}. We see:
\begin{itemize}
    \item a magical input box on the left,
    \item a form-based representation of the second sentence on the top right,
          which allows for simple editing of the constructor instantiation, and
    \item the output in two different languages on the bottom right.
\end{itemize}

The magical input box could indeed use a parser to make a suggestion in the form.
But this parser does not have to be high precision:
it is basically used to create a first suggestion that then
can be refined by the contributor.
A simple classifier will probably be sufficient, and can be easily
made available for a larger number of languages than a full-fledged parser.

\paragraph{The set of available constructors has to be extensible by the community.}
The constructors used to express the content and their individual keys and
whether these keys are required or optional, etc., have to be editable and extensible.
We cannot assume beforehand that we can create a full set of constructors that will allow us to capture Wikipedia.
This also means that the system has to be able to deal with changes of these constructors
--- we cannot assume that a constructor will be created perfectly on the first try.
When a constructor changes, content has to be transferred with as little loss as possible
(but we can assume that bots~\cite{bots1,bots2} can run over the content and make changes,
similar to the way they are used in Wikidata for clean-up tasks).
All changes to the constructors have to be reversible,
in order to allow for the wiki-principles~\cite{wikibook} that allowed Wikipedia to grow to remain applicable.

\paragraph{The renderers have to be written by the community.}
The only way to scale the creation of the renderers to hundreds of languages is
to enable the community to create and maintain them.
This does not mean that every single community member must be able to write renderers:
it will be a task which can only be done by contributors who dedicate some time to learn the syntax and the system.
The current Wikipedia communities have shown that contributors with very different skill sets can
successfully work together~\cite{wikipediaroles}:
some people gain deep knowledge in using templates in MediaWiki~\cite{mediawiki},
others write bots in Java~\cite{javabot} or Python~\cite{pythonbot} to operate on a large amount of articles,
others create JavaScript widgets~\cite{jsgadget} on top of Wikipedia,
and others contribute content through the VisualEditor, a WYSIWYG interface to Wikipedia~\cite{ve}.

\paragraph{Lexical knowledge must be easy to contribute.}
The renderers will have the task to translate constructors into natural language.
This will require large amounts of lexical knowledge, far larger than the renderers themselves.
Fortunately, Wikidata has been recently extended in order to be able to express and
maintain the lexical knowledge needed for the renderers~\cite{nielsen2019danish, nielsen2019ordia}.
This lexical knowledge base will be able to provide the necessary forms to render the content,
e.g. in the example above the German genitive case for Northern California,
the superlative for ranking financial centers by importance, etc.

\paragraph{Content must be easy to contribute.}
Just as lexical knowledge, content itself must be easy to create, refine, and change.
Content will continue to constitute the largest part of Wikipedia by far,
followed by the lexical knowledge and then, far behind, the renderers and the core software running the project.
If trade-offs in the designs of the different systems need to be made,
these should take into account how many contributors will be necessary for each of these parts.
The mockup in Figure~\ref{fig:mockup} suggests a system where we can edit a sentence through a form-based interface and
additionally have a parser for natural language that tries to guess the abstract syntax,
and allows the contributor to refine what they have entered in real-time.
This has the advantage that contributors are getting immediate feedback and
thus learning to express themselves in the constrained language the system understands.

\paragraph{All pieces are editable at all times.}
The system will not and cannot follow a waterfall process~\cite{waterfall}:
we will not see first the definition of all necessary constructors, then the creation of the required renderers,
then identify and provide the required lexical knowledge, and finally use that to write the content.
Instead, all of these parts ---as well as other parts of the infrastructure--- will change and improve incrementally.
The system must be able to cope with such continuous and incremental changes on all levels.
On the other hand, we do not have to assume that these changes are entirely arbitrary:
on Wikipedia we currently have many tens of thousands of templates and millions of invocations of these templates.
A change to a template might require the change to all articles using the template,
and often the template calls are found to be inconsistent and articles may be broken because of that.
The community has developed processes to notice and fix these.
It is OK to make certain assumptions with regards to the way the different parts of the system may evolve,
but they have to allow for the organic growth of all the pieces.

\paragraph{Graceful degradation.}
The different languages will grow independently from each other at different speeds.
Because different languages will have different amounts of activity,
some languages are expected to have very complete renderers, a full lexical knowledge base,
and to keep up-to-date with changes to the constructor definitions,
whereas other languages will have only basic lexical knowledge, incomplete renderers,
and  stale support for the existing constructors.
It is important that the system degrades gracefully, and doesn't stop rendering the whole article
because of a single missing lexicalization.
At the same time, the system must not behave entirely opportunistic.
There must be a way to mark dependencies between parts of the text,
so that some text requires other text to be rendered and
will be omitted if the requirement is missing,
even though the text itself could be rendered.

\section{Architecture}
\label{sec:architecture}

\begin{figure}[tb!]
\centering
\includegraphics[scale=0.6, trim=50 50 50 30, clip]{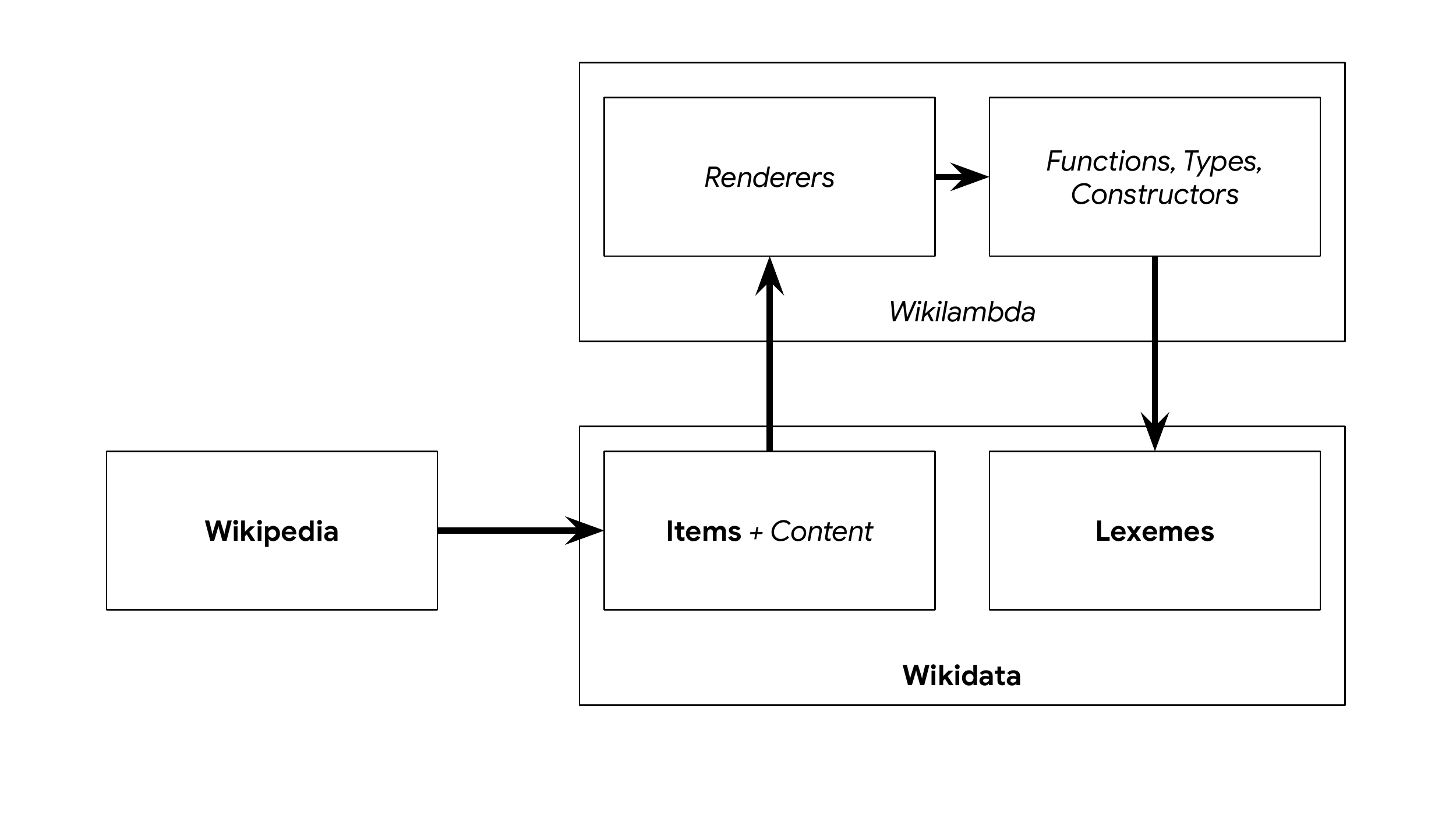}
\caption{
Architecture of the multilingual Wikipedia proposal.
To the left is the Wikipedia language edition.
The Wikipedia language edition calls the content stored in Wikidata
and the content is then rendered by renderers in Wikilambda.
The result of the rendering is displayed in Wikipedia.
The renderers are functions in Wikilambda, and there they can use other functions,
types, and constructors from Wikilambda.
These function finally can call the lexicographic knowlegde stored in Wikidata.
In italics are the parts that are proposed here (Content, Wikilambda),
in bold the parts that already exist (Wikipedia, Wikidata without Content).
}
\label{fig:architecture}
\end{figure}

The main point of dissemination of encyclopedic knowledge remains the
Wikipedia language editions~\cite{wikipedia1,wikipedia2}.
They can fetch content from Abstract Wikipedia, a new part of Wikidata~\cite{wikidata} alongside the items.
The content is than turned into natural language using the renderers stored in Wikilambda.
The resulting text is displayed in Wikipedia.
The renderers in Wikilambda are written using other functions, types,
and constructors stored in Wikilambda.
These functions can also call to the lexicographic knowledge stored in Wikidata~\cite{wikidatalex}.
This architecture is sketched out in Figure~\ref{fig:architecture} and
will be discussed in detail in the rest of this Section.

\subsection{Content in Abstract Wikipedia}

Abstract Wikipedia provides natural-language independent abstract content for the items in Wikipedia.
We propose to store this content in a new namespace
associated with the items in Wikidata~\cite{wikidata}.
Namespaces~\cite{mediawiki} are currently used e.g. for the talk pages in many wikis,
for citations in the English Wiktionary, etc.
Since the set of items in Wikidata has originally been seeded by
tying the Wikipedia articles in different languages together~\cite{wikidata},
they provide a natural list of topics that have Wikipedia articles.
The example constructor given in Figure~\ref{fig:example} could be
the start for the article on San Francisco in Abstract Wikipedia.

Abstract Wikipedia allows creating, extending, editing, and maintaining content.
An appropriate UX on top of it ---maybe as suggested in Figure~\ref{fig:mockup}---
would make sure that the content can be edited easily.
So in the unlikely event that San Francisco grows and surpasses the population of San Jose,
a contributor could just remove San Jose from the list on the field \texttt{after},
and change the value of the field \texttt{Rank} from \texttt{4} to \texttt{3}, store the change,
and see the text be updated in all language versions using the text.

In fact, there is some room for optimism that editing that kind of content might be more engaging
than editing prose on mobile devices,
as it might be supported by the interface modality of a mobile device more naturally than prose does.
But we recognize that getting this user interface right is both important and challenging.

Content is described by the simple grammar given in Figure~\ref{fig:grammar}.

\begin{figure}[tb!]
\centering
\begin{verbatim}
<content> ::= <constructor> ["(" [<argument> ["," <argument>]*] ")"]
<argument> ::= <key> ":" <value>
\end{verbatim}
\caption{Grammar for abstract content.}
\label{fig:grammar}
\end{figure}

Constructors are defined in Wikilambda.
For every constructor, there is a page in Wikilambda that explains the constructor,
what it is used for, the keys that constructor has, whether the keys are optional or not,
and what type of values are allowed for each key.
Constructors have a discussion page associated with them,
are editable, and created and maintained by the community.

A constructor instantiation is a constructor with its keys filled with values, similar to a function call,
but constructors are data structures.
The values can be literals
(such as the number \texttt{4} in the \texttt{Ranking} constructor on the key \texttt{rank}),
Wikidata items
(such as for the keys \texttt{subject} or \texttt{local\_constraint} in the same constructor instantiation),
lists (such as list of cities on the \texttt{after} key),
enumeration values (such as the conjuncts within the \texttt{Instantiation} constructor,
basically zero-key constructors that capture a certain semantic but do not have Wikidata items),
function calls (which first get executed and then replaced by their result),
or other constructor instantiations
(such as for the \texttt{class} key in the \texttt{Instantiation} constructor).
The constructor specification states which types of values are accepted for a given key.
The constructor specification also states the type of the result of the specification when being rendered,
i.e. if it is a noun phrase, a clause, or a string.
Most of the renderers are expected to return complex grammatical types, not strings.

Wikilambda hosts the constructor specifications, the specification and implementations of functions,
as well as the specification of possible types used for values.
Wikilambda can also host literal values for the types defined in Wikilambda,
thus enabling enumerations (as mentioned above, sometimes in lieu of proper Wikidata items) and other types with their values.

\subsection{Text in Wikipedia language editions}

Once the abstract content is stored, individual language Wikipedias can make calls
to the functions in Wikilambda and the content in Wikidata
(similarly to the way a Wikipedia article can already access data
from Wikidata~\cite{wikidataaccess}).
The results of these function calls are then displayed in the Wikipedia article.

Wikipedia (as well as other Wikimedia and external projects) call the \texttt{render} function
(see the following Subsection)
with some content and the language of the given Wikipedia as the arguments,
and display the result as article text to be included in the article and made available for the reader.
The article text would, on first glance, look very similar to a normal handwritten article on the given Wikipedia.
This way a Wikipedia, such as the Twi Wikipedia,
could call out to Wikilambda and Abstract Wikipedia and display the content rendered in
Twi.\footnote{Twi is the language of nine million native speakers, mostly in Ghana,
and yet they have fewer than ten active contributors.}

There are a number of simplifications that could be provided in order to improve
the coverage in Wikipedias with smaller communities,
and the local communities would decide which ones of those they would choose to activate.
For example, articles could be created
from Wikidata directly by adding a sitelink to an item that has none for the given language,
or we could even create all articles that generate results automatically.
We expect the different communities to autonomously decide how much of Abstract Wikipedia they want to use.

\subsection{Functions in Wikilambda}

The most relevant function in Wikilambda with regards to Abstract Wikipedia is
a function to render an abstract content in a natural language.
There will be a lot of work behind the following interface~\cite{dalereiter1,dalereiter2,dalereiterupdate}:

\texttt{render (content, language)} $\rightarrow$ \texttt{string}

In Wikilambda, functions have stable identifiers. A function interface can have several implementations.
Tests, pre- and postconditions are defined for functions.
Implementations of functions in Wikilambda can either be native in a programming language
such as JavaScript~\cite{js}, WebAssembly~\cite{webassembly},
or C++~\cite{cpp}, or can be composed from other functions.

A testing system can execute different implementations,
compare what the tests provided in Wikilambda say the results of a function should be,
and also compare the different implementations with each other
regarding their runtime characteristics and results.
Based on the analysis of the runtime characteristics,
an evaluator can choose an appropriate evaluation strategy
from the implementations available in Wikilambda.

Here is an example for an interface and an implementation via composition:

\texttt{multiply(positive\_integer x, positive\_integer y)} $\rightarrow$ \texttt{positive\_integer:}
\begin{verbatim}
  if(condition: is_zero(x),
     then: 0,
     else: add(y, multiply(subtract(x, 1), y)))
\end{verbatim}

For this example, \texttt{is\_zero} takes a \texttt{positive\_integer} and
returns \texttt{true} if it is \texttt{0} and else \texttt{false},
and \texttt{add} and \texttt{subtract} return the result of
the addition or natural subtraction\footnote{Natural subtraction
returns zero instead of a negative result, i.e. it always returns a natural number.}
of the two arguments, respectively.
This obviously will not lead to a particularly efficient evaluation of the function,
and, if available, an alternative implementation will likely be selected by the evaluator.
The type \texttt{positive\_integer} is an example of a type defined in Wikilambda,
and not meant to be a primitive type.

In order to evaluate a function call, an evaluator can choose from a multitude of backends:
it may evaluate the function call in the browser, in the cloud, on the servers of the Wikimedia Foundation,
on a distributed peer-to-peer evaluation platform,
or natively on the user's machine in a dedicated hosting runtime,
which could be a mobile app or a server on the user's computer.
Depending on the backend, some implementations might have very different run time characteristics:
a C++ implementation might be blazing fast in a local hosting runtime whereas
in the browser a JavaScript VM running GCC might be required, which would be a major cost on efficiency.
In the end, it is up to the system to measure or reason about the different options
and then decide dynamically which evaluation strategy to pursue.

Functions are, if not otherwise marked, truly functional--- i.e. they result in the same value given the same inputs,
which allows for effective caching strategies and optimizations for the evaluator.
This is particularly relevant for generating natural language, where Zipf's law tells us
that certain phrases and renderings are much more frequent than others~\cite{zipf}, and thus can benefit from caching.
Previous work has shown that a functional language is well suited
for a natural language generation task such as this one~\cite{gf}.

Function calls to Wikilambda can be embedded in many different contexts.
Wikilambda will provide UIs to call individual functions,
but besides that it also lends itself to a web-based REPL, locally installed CLIs, a RESTful API,
as a library which can be embedded in any programming language and thus making the functions universally accessible,
through dedicated apps installed on mobile devices, Jupyter notebooks, natural language assistants, spreadsheet cells,
or from a Wikipedia context, where function calls can be embedded,
most importantly to render contents from Wikidata in natural languages.

\subsection{Templates in Wikipedia}

Allowing the individual language Wikipedias to call Wikilambda has an addtional benefit.
It allows the different Wikipedia communities to unify their template code in a single place,
and make sure that their functions remain up to date.
Templates in MediaWiki~\cite{wikitemplates} are a widely used mechanism to unify parts of the content
of a given wiki by externalizing it in an external page, where other pages in the wiki
can call it.
Templates can be parameterized.
This makes template calls in Wikipedia similar to function calls.
But template interfaces are not well separated from their implementation,
an omission which is usually fixed by the community
diligently documenting and maintaining widely used templates.
Templates are used in Wikipedia for many different purposes,
from creating tabular overviews of sport tournaments to providing unit conversions or
calculating the age of a living person based on their date of birth.
Templates are implemented in wikitext~\cite{mediawiki} or Lua~\cite{lua}.

Templates are currently limited to a single wiki,
i.e. to an individual language edition of a project.
So improvements to a template made in the English Wikipedia do not
automatically propagate to the other language editions.
Nor do the contributors of the English Wikipedia consider the effect of changes to a template
on the other language editions if the changes were manually copied over.
This leads to the templates often diverging, which makes it harder
for contributors to bring content from one Wikipedia to another,
or even just to contribute in project discussions, where the knowledge and
command of the widely used templates of the given community,
such as \texttt{\{\{support\}\}}, \texttt{\{\{ping\}\}}, etc.,
make the contributors both more effective in their communication and act as social markers.
The templates often also contribute to the readability and attractiveness of a page.

By allowing the language editions to call functions in Wikilambda,
we allow the communities to relieve themselves of
the maintenance and creation of these templates if they are so inclined.
This again is nothing we expect to see in large communities
such as the English Wikipedia who numerous contributors with strong technical skills,
but it is expected to be helpful for smaller communities who prefer to focus
the activity of their contributors on their content
instead of technical aspects.
They would be able to call functions in Wikilambda,
and be able to benefit from the shared work of the more technical contributors there.

\subsection{Expressiveness}

Wikilambda allows for arbitrary functions, and thus it can be applied well beyond Abstract Wikipedia.
One question is: why should we allow for arbitrary functions?
Wouldn't we be able to simplify the system, have better guarantees on runtime,
and possibly increase the number of contributors,
because limiting what can be done in Wikilambda and by adding a set of constraints
can lead to a simpler user experience?

There are several problems with such limitations.
First, any set of constraints would likely be inspired by a strong commitment to a specific linguistic theory.
But looking at the state of natural language generation research,
there seems to be no consensus commiting to any specific linguistic theory.
Some systems are built on combinatory categorial grammars~\cite{ccg},
many systems on systemic-functional grammars~\cite{sfg, sfgnlg},
others on lexicalized tree adjoining grammars~\cite{ltag},
and many on head-driven phrase structure grammars~\cite{hpsg}, etc.
Since Abstract Wikipedia would stretch what most of the existing systems cover in terms of languages and domain,
it is likely that we do not know yet how to best approach this issue.
Therefore a guidance from theoretical assumptions to find interesting restrictions seems premature.

Second, even if some constraints were to be identified,
it is likely that these constraints would not reduce the resulting language to become less than Turing-complete.
Particularly if we allow for recursion, and almost all grammars do,
it is extremely difficult to avoid accidentally slipping into
Turing-completeness~\cite{turingmmcg, turinglisp, turingtmg, turingmov, turingc, turingcell}.
Instead of carefully avoiding it, the suggestion is to instead embrace it and plan accordingly.

Third, even if an effective set of constraints to reduce the expressivity of
the natural language generation language were available,
this would only apply for the strictly linguistic content of Abstract Wikipedia.
There are other forms of content: embedded multimedia, tables, graphs, infoboxes, etc.
which might be hampered by a linguistically motivated, less expressive language.

In summary, an architecture where Abstract Wikipedia runs on a Wikilambda that supports
the full expressivity of programming languages seems to be unavoidable anyway.
And given that, we should embrace it as an opportunity.
One of the core arguments of this proposal is that in order to build Abstract Wikipedia, we also have to build Wikilambda.
The following Section discusses further advantages of creating Wikilambda.

\section{Wikilambda}
\label{sec:wikilambda}

Imagine that everyone could easily calculate exponential models of growth,
or turn Islamic calendar dates into Gregorian calendar dates.
That everyone could pull census data and analyze the cities
of their province or state based on the data they are interested in.
That they could encode their local and historic units of measurements
and make that knowledge available to everyone in the world.
To analyze images, videos, and number series that need to stay private.
To allow everyone to answer complex questions that today would require coding and
a development environment and having a dedicated computer to run.
To provide access to such functionalities in every programming language,
no matter how small or obscure.
That is the promise of Wikilambda.

The main goal of Wikilambda is to allow for Abstract Wikipedia, as discussed above:
the abstract representation that is used to store the content of Abstract Wikipedia
is turned to natural language by using functions stored in Wikilambda.
This will lead to the creation of an open, widely usable,
and well tested natural language generation library on the Web, covering many different languages.
But besides Abstract Wikipedia, this also allows for a number of secondary goals
that will enable novel use cases and empower users and contributors all over the world
to participate more fully in the world of open knowledge.

Wikilambda will provide a comprehensive library of functions,
to allow everyone to create, maintain, and run functions.
This will make it easier for people without a programming background to reliably compute answers to many questions.

Wikilambda would also offer a place where scientists and analysts could create models together,
and share specifications, standards, or tests for functions.
Wikilambda provides a persistent identifier scheme for functions,
thus allowing you to refer to the functions from anywhere with a clear semantics.
Processes, scientific publications, and standards could refer unambiguously to a specific algorithm.

Also, the creation of new development environments or programming languages or paradigms will become easier,
as they could simply refer to Wikilambda for a vast library of functions.
This means that any new language could ``come with wheels included'' merely by linking to Wikilambda.

One obstacle in the democratization of programming has been that
almost every programming language requires first to learn some basic English.
Wikilambda will use the same schema as Wikidata to allow for an inherent multilingual usage.
It will be possible to read and write code in the native language of any of the contributors of Wikilambda,
as well through outside development environments that use Wikilambda as a repository of functions.

\section{Risks and open questions}
\label{sec:risks}

The proposal is ambitious and has a high possibility of not achieving its main goals fully.
In this Section we discuss the main risks that can stop the projects from achieving the whole breadth of the goals.

\begin{itemize}
    \item Leibniz was neither the first nor the last,
      but probably the most well-known proponent of a universal language,
      which he called \textit{Characteristica universalis}~\cite{characteristicauniversalis}.
      Umberto Eco wrote an entire book describing the many failures towards this goal~\cite{perfect}.
      Abstract Wikipedia is indeed firmly within this tradition,
      and in preparation for this project we studied numerous predecessors.
      The main difference between Abstract Wikipedia and Leibniz' program is that Leibniz not only aimed for
      a notation to express knowledge but also for a calculus
      to derive the veracity of its content and to answer questions.
      We solely focus on the part of expression,
      and only as far as needed to be able to render it into natural language,
      not to capture the knowledge in any deeper way. This gives the project a well defined frame.
      (We understand the temptation of Leibniz' desire to also create the calculus,
      and regard Abstract Wikipedia as a necessary step towards it, on which, in further research, the calculus
      ---i.e. a formal semantics to the representation--- can be added.)
    \item By creating a single Abstract Wikipedia the individual Wikipedia editions draw from,
      we create a single point of failure.
      Instead of having hundreds of independent Wikipedia editions,
      we now allow an attacker to corrupt a single place and get their bias spread throughout.
      (On the other side, the previous situation provides a deceptive sense of safety:
      in fact, it is easier to hijack a small community than a large one~\cite{wp20vrandecic}.
      We expect the Abstract Wikipedia community to become quickly larger
      than most individual language Wikipedia communities.
      Whereas it might be true that the hijacking of an individual language is contained to that language,
      it is also true that any speaker of that language does not benefit at all from
      the fact that there are hundreds of other languages out there which do not have that bias.)
    \item A gatekeeper mentality could take hold of Abstract Wikipedia
      which would not allow the community to grow to its full potential.
      Especially, because contributing to Abstract Wikipedia will possibly have
      a higher bar than contributing to most Wikipedia projects,
      this bar can further manifest in the Abstract Wikipedia community through additional
      social gatekeeping processes.
      This risk needs to be countered with an explicit effort accompanying the early growth of the community
      and a firm commitment to practical openness.
    \item Whereas Wikidata and Wikimedia Commons both exemplify communities
      with diverse contributors not sharing a single language,
      both projects are in a sense lucky as their content lends itself less to requiring
      detailed discussions and debates.
      Due to its structured and fact-oriented approach, Wikidata does not have
      as much potential for disputes as Wikipedia does.
      And the major discussions in Commons are about whether an asset is out of scope or not,
      and thus a binary decision.
      The more Abstract Wikipedia grows, the more it will lend itself
      to the same fierce debates that we see in the Wikipedias~\cite{editwars1,editwars2}:
      how much weight do we give to a certain point of view? How do we order the arguments?
      How do we express certain conflict-laden narratives?
      Given that unlike in the individual Wikipedia editions we explicitly must not assume
      that all contributors share a language, these debates can turn problematic.
      (Unless we capture the debates in an abstract form as well, but it is unclear how effective that can be
      and how soon we can reach such a level of self-hosting.)
   \item Having a single Abstract Wikipedia has the potential to significantly reduce knowledge diversity
      which is currently found in the wide array of individual language Wikipedia projects~\cite{omnipedia}.
      As we discussed before, this assumption is false.
      First, because we do not replace the current Wikipedia editions
      but merely allow them to add more knowledge from a central repository,
      none of the current knowledge diversity will be removed.
      We do not require any of the current editions to replace their local content.
      Second, the assumption that the current setup increases diversity effectively is false
      as most readers read a very limited number of languages~\cite{userlanguages},
      so the fact that other language editions provide other points of views is entirely irrelevant for them.
      In order to effectively increase the knowledge diversity of Wikipedia,
      each point of view needs to be reflected in all languages.
    \item Contributing to Abstract Wikipedia and Wikilambda may in many places require more expertise
      than contributing to many places in Wikipedia today.
      It is crucial to create a welcoming and inclusive user experience
      that allows a wide range of contributors to effectively contribute.
      Basically, we must make sure that it does not become too hard to contribute.
      We rest our hopes with the fact that already today, the most effective Wikipedia projects
      are those that manage to bring together a wide variety of contributors with complementary skills.
      Already today, tasks such as template editing, running bots,
      or writing Lua modules can only be performed by sufficiently skilled contributors.
      Many of the smaller communities are in fact incapable of nursing
      a sufficient number of contributors with the necessary diverse skill sets.
      The hope is that Abstract Wikipedia and Wikilambda will have a community that is diverse enough in this regard,
      and that at the same time will allow the local communities to outsource some of these necessary skills
      to Abstract Wikipedia and Wikilambda.
      This would allow the local communities to grow more effectively than they currently can.
    \item From a linguistic point of view, a major requirement for Abstract Wikipedia to achieve
      the goal of effectively generating content for many languages is that
      the number of constructors that need to be supported by a given language is not too high.
      We hope that the number of constructors is in the hundreds or low thousands,
      which would allow a community of five to ten active contributors to unlock
      a current and comprehensive encyclopedia for their language.
      Some preliminary work to estimate the number of necessary constructors using
      coverage checks based on FrameNet~\cite{framenet,framenet2,framenet3} allow us to estimate that
      about 2,500 constructors could provide promising coverage for encyclopedic texts.
      But should the number of constructors end up in the tens of thousands or higher,
      and if these are evenly distributed across the content,
      these hopes are not likely to realize.
      Results in natural semantic metalanguage~\cite{primes} even offer hope
      for a much smaller number of necessary constructors.
    \item One risk is that the approach outlined here
      ---Abstract Wikipedia content being rendered through Wikilambda functions for reading in Wikipedia---
      requires simply too much computing resources.
      It certainly will require significantly more computational power than the current approach.
      We hope that smart evaluation and caching strategies will manage
      to keep the necessary resources to a reasonable level.
      Particularly, caching is a very promising mechanism
      thanks to the functional nature of Wikilambda~\cite{functioncache1,functioncache2}
      and for Wikipedia in general,
      given the huge disparity between the number of readers and contributors,
      which has already been very helpful in keeping the cost of running Wikipedia low.
\end{itemize}

\section{Unique advantages}
\label{sec:advantages}

Whereas the project we sketch out is ambitious, there are several constraints we can take advantage of:
\begin{itemize}
    \item we aim only at a single genre of text, encyclopedias.
    \item in particular, we do not aim at representing poems, literature, dialogues, fiction, etc.
    \item fortunately, encyclopedic text has only
      locutionary content and no or little illocutionary force,
      meaning the system can focus on semantics without having to concern itself too much with pragmatics
    \item in general, the exact surface text that is rendered is not so important,
      as long as it contains the necessary content and, even more importantly, does not introduce falsehoods.
    \item in cases where the text cites other genres and relies on a high-fidelity reproduction of the text
      for a citation (e.g. when discussing Shakespeare citing \textit{``Shall I compare thee to a summer's day?''}),
      we use a mechanism to cite strings verbatim.
    \item it is fine to start with extremely simple sentences and allow the community to iterate over them.
      In fact, this is probably the only way we can grow the project.
      The system will likely never achieve the same expressivity as natural language,
      but already the ability to express simple statements and the possibility for the community to grow
      this ability over time is expected to make huge amounts of knowledge available to readers
      who previously could not access it,
      as can be seen with Lsjbot in Wikipedia language editions such as Swedish or Cebuano~\cite{lsjbot}
      ---and Lsjbot is not much more sophisticated than mail merge~\cite{mailmerge}
      regarding the natural language generation,
      capabilities which should be easily implementable in Wikilambda.
    \item we do not have to bother with parsing natural language, merely with generating it.
      This means we do not need to understand the entirety of a language,
      but need to merely be able to generate a small subset of possible utterances.
      Whereas it would be very helpful to have a parser or a classifier to help
      with the task of initially creating the abstract content,
      this is no blocking requirement as there are alternative user experiences
      for the contributor to create and maintain the content.
    \item with Wikipedia, we have the unique opportunity to rely on a large number of volunteer contributors.
      Whereas it is obvious that the system should be designed in such a way that it reduces
      the effort that is necessary to be performed by human contributors,
      it can ---and has to--- rely on human contributions and not on machine intelligence to solve all problems.
    \item finally, the baseline is very low. Below we list alternatives to this system,
      and their shortcomings are so major that even a modest success with achieving the proposed system
      will be a major improvement over the current situation.
\end{itemize}

\section{Opening future research}
\label{sec:open}

Wikilambda and Abstract Wikipedia are expected to catalyze work in a number of different
research areas.
Some of these lines of research are expected to bring direct benefits to the projects themselves,
but many should lead to results which will carry external benefits.
We see four main research areas that will benefit and be challenged by Wikilambda:
knowledge representation, natural language generation, collaborative systems,
and computer-aided software engineering.

\subsection{Knowledge representation}

The trend during the last two or three decades in knowledge representation has
been going toward understanding the limits of efficient and effective reasoning~\cite{dl,kr}.
Abstract Wikipedia will push towards a knowledge representation that is
closer to narrative and natural language.
Its success is to be measured by the question
\textit{``Can I render it in natural language?''}
and not \textit{``Can I derive inferences from it?''}.
How close can we get to an interlingua~\cite{unl,amr}?
This will lead to the desire of an algebra on top of Abstract Wikipedia,
i.e. how to add a semantics to the constructors that allow for inference,
consistency checking, and question answering.
The missing second part of Leibniz program~\cite{characteristicauniversalis}, basically,
which we explicitly declare as out of the initial scope for Abstract Wikipedia.

This could lead to a new format to capture the meaning of natural language that,
given the ready availability of Wikilambda, would be easy to share and evaluate.
This format can be used beyond Wikipedia,
for example for emails, in messaging apps, by digital assistants,
allowing for high-fidelity communication across languages.
Genres for which this is particularly interesting are
legal~\cite{legal}, health~\cite{health}, and scientific texts~\cite{science},
as they could make such texts easier to check, standardize,
summarize, and present in context.
This format can improve information extraction considerably~\cite{templateie},
as it does not have to deal with issues of
the surface text of individual natural languages anymore. 

\subsection{Natural language generation}

Abstract content can be used to personalize the text beyond
just the choice of natural language.
Text can be created for different reading levels~\cite{readinglevel}.
It can also take into account previous knowledge by the reader.
It can be made more age-appropriate, be more concise for for expert readers,
or take into consideration the effect of certain disabilities.
It will be simpler to select the amount of detail in the text,
to accommodate both the hurried reader and the curious student~\cite{zoom}.
Abstract content can be a starting point toward generating other modalities,
such as adapting it to being spoken~\cite{speech} or accompanying video.

A local Wikipedia might intentionally decide to divert
from the output by Abstract Wikipedia,
which leads to an easier evaluation of the knowledge diversity of Wikipedia.
This can help in uncovering and dealing with biases
in less active Wikipedia editions~\cite{omnipedia}.
But we also must make sure that we don't just enable large communities
to spread their biases now more globally and effectively.

Wikilambda can be used to create high quality
parallel or mono-lingual corpora in many different languages,
which in turn can be used for training machine translation and
natural language processing systems particularly for resource-poor languages~\cite{lowresource,zeroshot}.
But when renderers are changed,
we need to ensure that they do not break previous results.
How do we evaluate such changes?
Also, can we learn renderers automatically,
either unsupervised from a text corpus or
in an active learning approach with contributors?

\subsection{Collaborative systems}

The project will require contributors with very different skill sets,
interests, and time constraints to work together.
How can we ensure that the community can self-organize such diverse contributors
so that their contributions will be most effective~\cite{wikipediaroles,wikipedia1,shirky}?
Can we show them gaps and the impact filling these gaps would have?
The impact of their previous contributions, in order to motivate them further?
How will contributors with different language backgrounds work together?
Can they use the abstract format to resolve their differences?
Bots and humans will be working together on a very different level,
given that bots can be used to edit, check, and maintain the content easier.
Can we tie in a simple feedback mechanism for more casual contributors?
How do we ensure that readers can easily become contributors? 

Machine learning can be used to classify natural language input and support the creation of content.
The same classifiers and parsers can run on the content of Wikipedia and other sources,
such as scientific articles or reference texts,
to detect and remedy knowledge gaps~\cite{gapfinder}.

\subsection{Computer-aided software engineering}

Wikilambda can catalyze research and development in democratizing access
to code and code development.
It is seen as an impediment that in order to learn to program,
people need to learn basic English first~\cite{englishrequired}.
Wikilambda and it's usage of language independent identifiers
(in the same vein as Wikidata) could lift that requirement.
Wikilambda also will reach new surfaces and allow more people to participate
in the development of code and also reaping the fruits of such developments.

The metadata, tests, and specifications of functions in Wikilambda
will constitute a large catalog that can be used to generate new implementations
by means of program synthesis~\cite{programsynthesis1,programsynthesis2},
machine learning~\cite{neurosynthesis}, constraints~\cite{constraints},
or evolutionary programming~\cite{evolution}.
These can be automatically compared to other manually provided implementations
in terms of speed, parallelizability, memory usage and other run-time behavior. 

We can also analyze existing local code bases to find pieces of code
that can be improved by faster implementations, or to discover subtle bugs.
Imagine a system that automatically improves an existing code base using the Wikilambda catalog.
Further imagine that IDEs are extended to automatically pull implementations from Wikilambda,
speeding up development time, ensuring that the most efficient and bug-free implementation is being used.
Can this push the programming interface to a new level of abstraction?
Wikilambda can catalyze the creation of new programming languages.
Instead of having to support a rich library providing many capabilities,
new programming languages can now simply provide an interface to Wikilambda.
This should help with getting new ideas started and spread.
Can we even aim to map natural language questions to Wikilambda function calls
and their compositions, thus allowing us to code via natural language?

The development of evaluation environments creates an interesting ecosystem,
particularly for cloud providers.
They can provide access to functions defined in Wikilambda,
without necessarily using the implementation from Wikilambda.
Due to the interface specification it would be easy to switch different providers,
who could then compete on providing efficient implementations.
In fact, cloud providers could even hide the actual costs and just compete in price,
thus reaping benefits from being more efficient than the competitors.
But even outside the cloud, how do we create efficient evaluation environments?
How to develop evaluation strategies for Wikilambda functions,
including smart caches, that can increase the efficiency of executing function calls?
How to integrate non-functional function calls effectively,
such as a random number generator,
asking for the current time, or calls to sensors~\cite{random}?

\section{Alternative approaches}
\label{sec:alternatives}

In this Section we take a step back and focus solely on the question whether
we can effectively close the knowledge gap for readers in many languages by other means.
This means that we specifically ignore the other secondary goals and progress
towards answering certain research questions
that we hope to achieve with Abstract Wikipedia and Wikilambda.

\subsection{Organic growth}
\label{ssec:organic}

The most extensively used approach is to try to grow the communities for each language independently,
and hope that for all languages we will eventually find enough volunteers
to create and maintain a comprehensive encyclopedia.
This has worked in a small number of languages,
and many of the projects and quite a large amount of funding is geared towards
the goal of strengthening the individual language communities and
organically grow the smaller Wikipedia language editions~\cite{organic1,organic2}.

To put the challenge into perspective: currently English Wikipedia has about 30,000 active editors.
If all existing language editions were to have the same number of active editors,
there would be around nine million active editors
---currently, there are about 70,000, so that would mean an increase by more than two orders of magnitude.
Also, a number of languages Wikipedia is available in do not have 30,000 speakers:
Manx, Mirandese, or Lower Sorbian are examples of languages with
fewer speakers than English Wikipedia has active editors.

In short, it is not only unlikely but partially impossible
to achieve Wikipedia's vision through organic growth.
Also, it has been tried for two decades, and it is time to consider alternatives.

There is also a case to be made that the Abstract Wikipedia will allow
for a new incentive infrastructure that will in fact increase
the number of contributors to all of the individual language editions,
particularly the smaller ones.
Because the Abstract Wikipedia will provide a realistic way to achieve the goal of
a comprehensive and up-to-date encyclopedia in a given language,
it is likely that contributors who currently don't believe
that goal is achievable will be more motivated to contribute,
thus leading to an increase of organic growth.

\subsection{Machine translation}
\label{ssec:mt}

Another widely used approach ---mostly for readers, much less for contributors--- is the use
of automatic translation services like Google Translate.
A reader finds an article they are interested in and then asks the service
to translate it into a language they understand.
Google Translate currently supports about a hundred languages --- about a
third of the languages Wikipedia supports.
Also the quality of these translations can vary widely --- and
almost never achieves the quality a reader expects
from an encyclopedia~\cite{googletranslate,translatecompare}.

Unfortunately, the quality of the translations often correlates with
the availability of content in the given language~\cite{lowresource},
which leads to a Matthew effect: languages that already have larger amounts
of content also feature better results in translation.
This is an inherent problem with the way Machine Translation is currently trained, using large corpora.
Whereas further breakthroughs in Machine Translation are expected~\cite{zeroshot}, these are hard to plan for.

In short, relying on Machine Translation may delay the achievement
of the Wikipedia mission by a rather unpredictable time frame.

One advantage Abstract Wikipedia would lead to is that Machine Translation system can use
the natural language generation system available in Wikilambda to generate high-quality and high-fidelity parallel corpora for even more languages,
which can be used to train Machine Translation systems which then can resolve the brittleness a symbolic system will undoubtedly encounter.
So Abstract Wikipedia will increase the speed Machine Translation will become better and cover more languages in.

\subsection{Content translation framework}
\label{ctf}

A promising approach is the Content translation framework developed by Wikimedia for
contributors inside Wikipedia~\cite{contenttranslationframework}.
This allows community members to take an article from one Wikipedia language edition,
and use it to start an article in another language edition.
An automatic translation service may be configured to help
with the initial text for the article for certain language pairs,
which then can be corrected and refined by the editor. The framework is adapted to wikitext.
The tool has already led to half a million articles being translated.

Content translation framework is a great success.
Its shortcomings are in the independence of the articles once they are translated,
thus not having an option to propagate updates and corrections through the different language editions,
and in the fact that machine translation only supports limited language pairs.
Also, the amount of work needed is still orders of magnitude higher
than the work Abstract Wikipedia will require for the same results.

The big advantage of the Content translation framework is that it fits in the current system with much less disruption,
and it already has proven to provide effective means to increase the number of articles in many languages.

\subsection{Other machine learning}
\label{ssec:ml}

The last decade has seen a massive increase in capabilities for machine learning models~\cite{nmt1,nmt2}.
But Abstract Wikipedia and Wikilambda are firmly rooted in classic symbolic approaches.

The reason is simple:
Abstract Wikipedia's foremost approach is in the tradition of Wikipedia, being built by a community of volunteers.
This requires that volunteers can perform effective edits,
edits where they can see the results immediately and
where they can understand how their contributions will change the project.
With machine learning, we do not know how to allow a wide demographic
to edit the models in such a predictable and effective way.
Even just understanding the results of machine learned model is in fact hotly researched question~\cite{xai1,xai2}.

We expect that machine learning approaches will become part of Abstract Wikipedia and Wikilambda.
There is nothing in Wikilambda that requires the implementation of a function to be classical symbolic,
it can just as well consult a machine learned model.
The main difference is that machine learned models often benefit
from taking a step back and tackling a problem more end-to-end,
instead of using modular approaches where they lose potentially crucial information at
the information poor interfaces between the modules.
So we expect that Wikilambda would need to provide sufficiently high-level functions
to have machine learned models to provide benefits being implementations.

Three examples of how machine learned models can be smoothly integrated into Abstract Wikipedia are the following:

\begin{itemize}
    \item \textbf{Overgenerate and rank.}
      Natural language often has a large number of possible ways to express a certain piece of information,
      and it can be very hard to encode how to choose between similar options.
      One way to turn this into an advantage and at the same time potentially
      improve the fluency and naturalness of the text is to encode different options in Wikilambda,
      to generate all of them, and then let a machine learned model select
      the best of these options~\cite{overgenerateandrank1,overgenerateandrank2}.
      This has the advantage that any of the generated options are still entirely grounded in the abstract content,
      and thus known to be correct, and we don't have to deal with
      the usual problem of machine learned natural language generation models such
      as hallucinating or dropping important grammatical details such as negation.
    \item \textbf{Classify natural language input.}
      It is likely that entering natural language is easier than using a form-based approach
      towards generating the initial content for an article in Abstract Wikipedia.
      In order to support contributions through natural language we need to have classifiers for these texts.
      Since these classifiers are only used during the creation of the text, machine learned classifiers can be used,
      as the result will be immediately checked by the contributor for whether the classification was successful,
      and the actual storage of the content would happen in the abstract format.
      Any errors from the classification wouldn't leak to the actual content,
      the classifier would solely be used to speed up the original creation of the abstract content.
      This would allows us to use state of the art language models for classification~\cite{bert}.
    \item \textbf{Refine results using autocorrect.}
      Since creating the renderers might be a large amount of work, it might be worthwhile to
      create simpler renderers that create the basic skeleton of the text, and then use a 
      system to autocorrect the results, improve the fluency of the text, and ensure that
      agreement and other grammatical constraints are being followed~\cite{lasertagger}.
\end{itemize}

A completely different approach based on a machine translation system would be to allow for
parallel authoring in several languages,
displaying translations in other languages of the contributor's choice~\cite{userlanguages},
as well as translations back to the original language.
The models for machine translation would be specifically trained to converge when being translated back and forth,
and the system would store all content in several languages at the same time.

Contributors could correct any of the supported languages, and could add more content in those languages,
which would, in turn, be used to train and improve the models.
It is unclear how improved models would be deployed,
given that it might perturbate the whole system with its parallel translations.
For new languages that the contributor does not speak, we can then use
multi-source translation systems using all of the languages together~\cite{multisource1,multisource2}.

Such a system would have a much easier usability for contributors than Abstract Wikipedia,
and would massively lower the bar to contribute, but it is unclear whether it would actually work,
and designing the user experience just right might prove challenging.

\subsection{Wikidata}
\label{ssec:wikidata}

Wikidata is a structured multilingual knowledge base~\cite{wikidata}.
The content of Wikidata is available to all Wikipedia languages (given that the labels are translated),
but compared to natural language the expressivity of Wikidata is extremely limited.
It is good to keep simple statements, mostly the kind of knowledge available
in the so-called Wikipedia infoboxes which are on the right hand side of many articles:
Berlin is located in Germany, Douglas Adams was born in Cambridge, Moby Dick was written by Herman Neville.
But it could not express how photosynthesis works, the reasons for World War II, or Kant's categorical imperative.
Wikidata is fast growing, and has more than a billion statements and an active community of volunteers.

Extending Wikidata's expressivity to capture a much larger amount of Wikipedia's content
would be extremely challenging and stretch the boundaries of knowledge representation far beyond the state of the art.
Once this is achieved it is unclear whether the user interface could actually support that expressivity.

Another issue is that Wikidata is centered on a single subject per page,
whereas most Wikipedia articles mention many other subjects while describing the main subject of an article.
Furthermore, a lot of the content of Wikipedia is redundant,
and this redundancy is essential for the ease of human understanding of the content:
Wikidata offers the population of San Francisco,
but it doesn't explicitly state that it is the fourth-largest city in California.
This would need to be deduced from the knowledge base as a whole.

From any decent sized knowledge base, it is trivial to infer an almost infinite number of additional facts.
But which of these facts are interesting? In what order should they be told?
How do we express the reasons for World War II or how photosynthesis works in a knowledge base like Wikidata?
It seems that turning a declarative knowledge base into good text is much harder than rendering
an abstract syntax (although it has been done in some limited domains with success~\cite{wikidatabios,summary}).

Wikidata, on the other side, can be used as a background knowledge base for Abstract Wikipedia.
It remains true that editing and updating, e.g. a population number or a mayor,
will be easier in Wikidata than in Abstract Wikipedia (particularly using bots, e.g. after a census).
Abstract Wikipedia can query Wikidata using SPARQL and format the results~\cite{sparql,sparqlwikidata}.

Wikidata's main shortcomings in capturing the expressive content we meet with Abstract Wikipedia are:

\begin{enumerate}
    \item relations that connect more than just two participants with heterogeneous roles.
    \item composition of items on the fly from values and other items.
    \item expressing knowledge about arbitrary subjects, not just the topic of the page.
    \item ordering content, to be able to represent a narrative structure.
    \item expressing redundant information.
\end{enumerate}

\subsection{Open Source}
\label{ssec:oss}

Another way to split the work would be to have the abstract content stored in Wikidata,
as suggested, but to not have Wikilambda at all and instead have the code be a proper part of the Wikidata code base.
That code base is then managed following the usual open source process~\cite{opensource1,opensource2}.

There are several disadvantages and advantages to this solution.
The major advantage is the familiarity with the process, and the available tooling to support this approach:
test frameworks, source control, IDEs, review processes, etc.
We have been doing distributed and open source development for decades, why not build on that experience?

The main disadvantage is that the common open source process comes with a large barrier to entry:
it requires to set up a development environment, to have sufficient access to computing power to run it, etc.
We hope that by surfacing the code to Wikilambda, in a manner that can be easily experimented with through the browser,
we can actually lower the barrier sufficiently to get people to contribute code who have never coded before.

In the end, Wikilambda aims to generate NLG templates for hundreds of languages,
which means we need thousands of contributors
---a small group for each of the languages.
Without a system that is geared towards reaching such a wide contributor base,
that aims at allowing people to easily contribute a bit of knowledge,
it is unlikely for Abstract Wikipedia to gather sufficient support.

Also, we expect the natural language generation code to simply be extensive.
This makes it more suitable to be part of the data editable through a wiki-based interface than code editable through a version control system.

\section{Summary}
\label{sec:summary}

Achieving a multilingual Wikipedia is a clearly defined goal with many challenging problems along the way.
With Abstract Wikipedia and Wikilambda, we sketch out a possible architecture to achieve that goal.
We have outlined risks, desiderata and unique advantages.

A major advantage of splitting up Wikilambda and Abstract Wikipedia is
that we acknowledge that there are major risks in the project,
but particularly regarding Abstract Wikipedia.
Wikilambda offers a number of valuable goals by itself, and Wikilambda has a much lower risk than Abstract Wikipedia.
Therefore with Wikilambda we identify a milestone that will achieve interesting results by itself.
And Abstract Wikipedia can provide value covering a lot of the current bot-created articles
and putting them into a more maintainable context.
So even if the full vision as laid out here is not achieved, there are many valuable intermediate goals.

A unique hope associated with this project is that we will expand widely who can contribute to today's most important knowledge source, Wikipedia.
Today, if a contributor doesn't speak English well enough, they are effectively unable to contribute to the English Wikipedia,
which is de facto the most important source of knowledge for many advanced technologies.
A machine learned approach by a huge company requiring massive resources,
based on the content of the whole Web and other content
will for a long time not become as open and as editable.
Abstract Wikipedia might be our best bet to ensure that each and every single one of us
retains the opportunity to contribute to the world's knowledge.

When discussing this problem with others, it is often mentioned how ambitious and hard the problem is
---which is particularly surprising when these people work on issues like artificial intelligence or machine translation.
It seems that the challenges inherent in AI and MT are not only necessarily harder than the challenges
we would encounter when realizing a multilingual Wikipedia, but they would be a true superset:
everything we need to realize a multilingual Wikipedia will also be needed for AI.

The project described here is achievable without the need for a major breakthrough in our current knowledge.
The current state of the art in natural language generation, lexical knowledge representation,
and collaborative systems can work together to create a novel system
that will enable everyone to share in the sum of all knowledge.

\section*{Acknowledgments}

\begin{flushleft}
Special thanks and love to Kamara Vrande\v{c}i\'{c} for her patience over the years, for her invaluable input,
her knowledge and intuition around languages, and her love and humor, which have shaped the ideas presented here.
I thank
\mbox{Enrique}~\mbox{Alfonseca},
Eric~\mbox{Altendorf},
Tania~Bedrax-Weiss,
Abraham~Bernstein,
Jiang~Bian,
Olivier~Bousquet,
Dan~\mbox{Brickley},
Bruno~Cartoni,
Nancy~Chang,
Brian~Chin,
Philipp~Cimiano,
Neboj\v{s}a~\'{C}iri\'{c},
Robert~Dale,
Jeroen~De~Dauw,
Michael~Ellsworth,
Colin~H.~Evans,
Charles~\mbox{Fillmore},
Tim~Finin,
R.V.~Guha,
\mbox{Macduff}~Hughes,
David~Huynh,
\mbox{Lucie}~Aim\'{e}e~\mbox{Kaffee},
Benedikt~\mbox{K\"{a}mpgen},
David~Karger,
Daniel~\mbox{Keysers},
Daniel~\mbox{Kinzler},
Yaron~Koren,
Markus~Kr\"{o}tzsch,
Tobias~Kuhn,
Stephen~\mbox{LaPorte},
Alex~\mbox{Lascarides},
Micha\l{}~\mbox{\L{}azowik},
Russell~\mbox{Lee-Goldman},
\mbox{Tatiana}~\mbox{Libman},
Erin~Van~Liemt,
Daphne~Luong,
Alex~MacBride,
Gordon~Mackenzie,
\mbox{Gerard}~\mbox{Meijssen},
Nikola~\mbox{Momchev},
Toby~\mbox{Negrin},
Natasha~Noy,
Praveen~\mbox{Paritosh},
Marius~Pasca,
Fernando~Pereira,
\mbox{H\'{e}ctor}~P\'{e}rez~Urbina,
Jo\~{a}o~\mbox{Alexandre}~\mbox{Peschanski},
Lydia~Pintscher,
Mukund~Raghavachari,
\mbox{Susanne}~\mbox{Riehemann},
\mbox{Jason}~Riesa,
Stuart~Robinson,
\mbox{Mariano}~\mbox{Rodr\'{i}guez~Muro},
Scott~Roy,
Daniel~Russell,
Nathan~Scales,
Silke~Scheible,
Nathanael~Sch\"{a}rli,
Hadar~Shemtov,
Max~Sills,
Elena~Simperl,
Javier~Snaider,
Rudi~Studer,
Vicki~Tardif,
Jamie~Taylor,
Barak~Turovsky,
Brion~Vibber,
Chris~Welty,
William~Woods,
Mickey~Wurts,
Leila~Zia,
and many others,
who, in one way or another, have shaped the ideas expressed here, and helped them grow,
sometimes without realizing it, but always humoring my obsessions with it.
\end{flushleft}

\bibliographystyle{plain}
\bibliography{ms}
\end{document}